\documentclass[aps,psfig,showpacs]{revtex4}

\usepackage{graphicx}
\usepackage{epstopdf} 
\usepackage{epsfig}
\input {epsf.sty}

\begin{document}

\draft
\title{Quantum Hall physics in rotating Bose-Einstein condensates}
\author{Susanne Viefers}
\address{Department of Physics,  University of Oslo,
P.O. Box 1048 Blindern, N-0316 Oslo, Norway }
\date{\today}
\begin{abstract}
The close theoretical
analogy between the physics of rapidly rotating atomic Bose condensates 
and the quantum Hall effect ({\em i.e.} a two dimensional electron gas in a strong
magnetic field) was first pointed out ten years ago. As a consequence of this analogy, a large number of strongly 
correlated quantum Hall-type states have been predicted to occur in rotating 
Bose systems, and suggestions have been made how to manipulate
and observe their fractional quasiparticle excitations.
Due to a very rapid development in experimental
techniques over the past years, experiments on BEC now appear to be
close to reaching the quantum Hall regime. This paper reviews the theoretical
and experimental work done to date in exploring quantum Hall physics in cold
bosonic gases. Future perspectives are discussed briefly, in particular the idea of
exploiting some of these strongly correlated states in the context of topological quantum
computing.

\bigskip
\noindent
{\it To appear as a 'Topical Review' in Journal of Physics: Condensed Matter}

\end{abstract}
\maketitle
\vspace{-11pt}

\eject

\tableofcontents
\eject

\newcommand{\half}{\frac 1 2 }
\newcommand{\eg}{{\em e.g.} }
\newcommand{\ie}{{\em i.e.} }
\newcommand{\etc} {{\em etc.}}

\newcommand{\noi}{\noindent}
\newcommand{\etal}{{\em et al.\ }}
\newcommand{\cf}{{\em cf. }}

\newcommand{\dd}[2]{{\rmd{#1}\over\rmd{#2}}}
\newcommand{\pdd}[2]{{\partial{#1}\over\partial{#2}}}
\newcommand{\pa}[1]{\partial_{#1}}
\newcommand{\pref}[1]{(\ref{#1})}

\newcommand{\ee}{\end{eqnarray}}
\newcommand{\e}{\varepsilon} 
\newcommand{\D}{\partial}
\newcommand{\pt}{\tilde p}
\newcommand{\p}{\partial}
\newcommand{\jas}{\prod_{i<j}(z_i - z_j)}



\newcommand {\be}[1]{
     \begin{eqnarray} \mbox{$\label{#1}$}  }

\section{Introduction}
Almost a decade ago, it was realized\cite{wilkin1} that there exists an intimate theoretical connection between two
seemingly very different physical systems: The quantum Hall effect(QHE)\cite{fqhe}, which occurs in a
two-dimensional electron gas (2DEG) at very low temperatures and strong magnetic fields,
and a rapidly rotating, dilute Bose-Einstein condensate\cite{smith+pethick} of electrically 
neutral atoms. The key to this analogy is the observation that in two dimensions, rotation
and a perpendicular magnetic field play a very similar role, making the two systems
mathematically equivalent. This implies that, at sufficiently fast rotation, a Bose condensate
is expected to enter a regime with strongly correlated states of the quantum Hall type, including
quasiparticle excitations that obey fractional (anyon) statistics\cite{leinaas77}. 

In recent years, the quantum Hall - rotating BEC
analogy has been theoretically explored in great detail, starting with the prediction\cite{cooper1} of a
bosonic Laughlin state\cite{laughlin83}. In addition, a large number of other incompressible states
are, in principle, expected to occur, including the bosonic analog of the Jain sequence\cite{jainreview}
and other Abelian states, but also non-Abelian states\cite{MR,RR}. Moreover, a number of 
exotic states have been predicted in the case of rotating bosons with spin\cite{reijnders}.

Unfortunately, experiments have not yet reached this quantum Hall regime. However, there
is reason to be optimistic, as the experimental development over the past decade has
been astounding. The first experiments on rotating atomic Bose condensates were performed
in the late nineties, and the first observation of a quantized vortex reported in 1999\cite{matthews99,madison00}.
Since then, Abrikosov lattices with hundreds of vortices have been produced, and 
present-day experiments\cite{schweickhard1}
are close to the rotation speed at which this vortex lattice is predicted to melt
and the system would enter the quantum Hall regime. The main obstacle is that at these
rotation speeds, the system is close to the point where the centrifugal potential cancels the
external harmonic trap and the atomic cloud would fly apart. There are, however, recent proposals
how to get around this problem by modifying the confining potential, so that the quantum Hall regime
may be reachable in the near future.

In addition to being interesting in its own right, the prospect of producing quantum Hall-type states
in cold atom systems may have long-term practical applications. One of the reasons
for the recently revived interest in the anyonic excitations of the QHE is the theoretical proposal to 
use them in the context of quantum computing\cite{topQCreview}. This vision is certainly
very far into the future. On the other hand, atomic systems may eventually turn out to be
superior to the 2DEG quantum Hall system, as they allow for a very large degree of controlled
tunability of various experimental parameters, including interaction strength and
details of the confining potential. Moreover, these systems are well isolated and clean
and thus less prone to decoherence than solid state realizations.

This paper presents a (hopefully) comprehensive review of this, still active, field of research.
Section \ref{sec:qhe} gives a brief overview of the fractional quantum Hall effect. In Sec.\ref{sec:RBEC}
we give a general introduction to the subject of rotating Bose condensates, summarizing the 
experimental developments of the past decade and the theoretical understanding of how the
system goes from a vortex lattice to the quantum Hall regime as rotation is increased.
Sec.\ref{sec:LLL} explains the theoretical equivalence between a fast-rotating Bose gas and
electrons in a strong magnetic field, along with some
of the basic properties of the resulting many-body spectrum in the presence of interactions.
An account of the literature on Abelian bosonic quantum Hall states is given in Sec.\ref{sec:aqhe};
most of this work is numerical or based on trial wave functions such as those of the composite
fermion phenomenology\cite{jainreview} and involves testing for the occurrence and stability
of incompressible states (and their fractional excitations) at, \eg the Jain fractions. 
This section also contains a brief discussion of the applicability of the composite fermion scheme
for very low angular momentum states.
Quite some work has been done to study the possible occurrence of non-Abelian quantum Hall
states, which are particularly interesting in the context of topological quantum computing. This
is accounted for in Sec.\ref{sec:NA}, which concludes that such
states appear to be more prominent in rotating BEC than in the conventional QHE. Following this,  
we summarize various recent proposals how to design experiments capable
of reaching the quantum Hall regime with present-day experimental techniques (Sec.\ref{sec:beyond} ).
Finally, we briefly discuss multicomponent Bose condensates (Sec.\ref{sec:spin})
and round off with come concluding remarks and future perspectives in Sec.\ref{sec:concl}.

\section{The fractional quantum Hall effect -- a brief overview}
\label{sec:qhe}
The fractional quantum Hall effect (FQHE)\cite{fqhe} is one of the most intriguing and most studied phenonema in condensed matter physics 
during the past 2-3 decades. It occurs in two-dimensional, high-mobility electron systems (typically formed at the interface between two 
semiconductor crystals, \eg in GaAs heterostructures) subjected to a strong magnetic field and low temperatures
(in the millikelvin regime in present-day experiments). 
Figure \ref{fig1} shows
a sketch of a typical Hall experiment: A current $I_x$ is passed through the sample along the $x$-direction and the resulting
transverse voltage $V_y$ measured for varying values of the magnetic field. Roughly speaking, the occurrence
of a transverse voltage can be understood as being due to the deflection of the charge carriers in the presence of the
external magnetic field, causing a build-up, or imbalance, of charge along the edges of the sample. In a purely classical picture,
this leads to a linear relation for the Hall resistance $R_{xy} = V_y / I_x$ as function of the magnetic field $B$.
\begin{figure}
{\psfig{figure=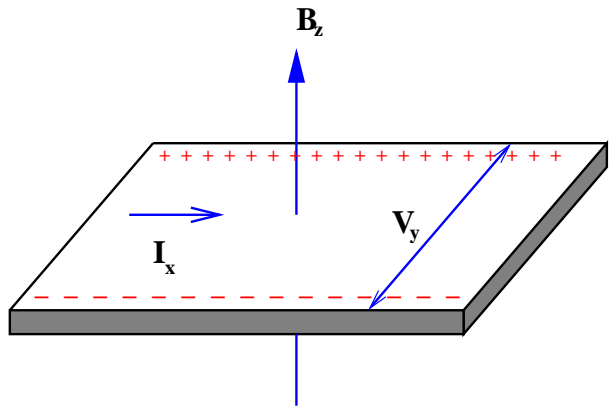, scale=0.5,angle=-0}}
\caption{Sketch of the Hall experiment. The 2DEG is exposed to a strong perpendicular magnetic field $B_z$.
A current $I_x$ is passed through the sample along the $x$-direction, and the resulting
transverse voltage $V_y$ measured for varying values of the magnetic field.}
\label{fig1}
\end{figure}
In the {\it quantum} Hall effect, however, the Hall resistance is quantized,
\be{Rxy}
R_{xy} = \frac{V_y}{I_x} = \frac{1}{\nu} \, \frac{h}{e^2},
\ee
where $h$ is Planck's constant and $e$ the electron charge. The number $\nu$ takes integer
values (integer quantum Hall effect) or equals rational fractions (fractional quantum Hall effect),
and each allowed value of $R_{xy}$ remains constant for a finite range of the magnetic field, as 
indicated in figure \ref{fig2}. At the same time, the longitudinal resistivity $\rho_{xx}$ equals
zero, except for the transition regions between neighbouring plateaux. A third characteristic
property of each of these quantum Hall states is that the system is {\it incompressible}, \ie
there is a gap between the ground state and (bulk) excitations. 
The integer effect was discovered in 1980 by von Klitzing et al\cite{klitzing}; two years later,
using even cleaner samples, Tsui and collaborators reported the discovery of the
fractional effect\cite{tsui82} at $\nu = 1/3$. Since then, with the fabrication of ever-higher mobility
samples, a large number of fractions have been observed\cite{disorderfootnote}. 
The quantization of the Hall resistance turned out to be extremely exact 
(to at least ten parts in a billion), which has led to the introduction of  a new standard of resistance, 
with the so-called von Klitzing constant 
$R_K = h/e^2$, roughly equal to 25812.8 ohms, as the fundamental unit.

\begin{figure}
{\psfig{figure=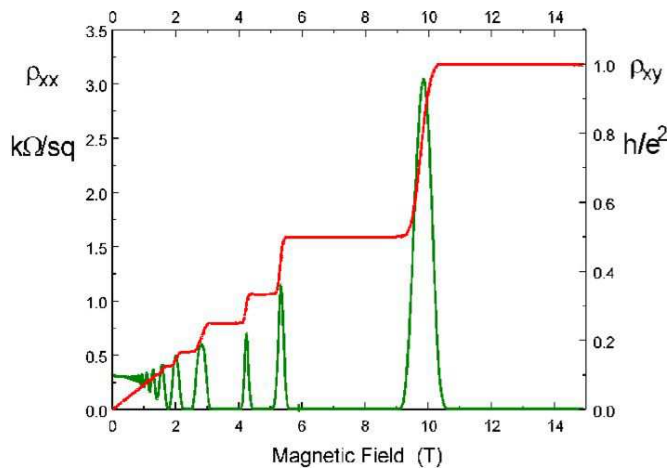, scale=0.35,angle=-0}}
\caption{Sketch of the (integer) quantum Hall effect. The Hall resistance as function of the magnetic field is
quantized, \ie exhibits plateaux. (The result predicted by the {\it classical} Hall effect corresponds
to a straight line through the centres of the plateaux.) The longitudinal resistivity  is zero except
at transitions between plateaux.
Courtesy of D.R. Leadley, Warwick University 1997.}
\label{fig2}
\end{figure}
Physically, the number $\nu$ in (\ref{Rxy}) corresponds to the Landau level filling fraction 
at the center of the corresponding plateau\cite{footnote1}. In other words, the IQHE occurs when
an integer number of Landau levels is filled, while the FQHE is seen at fractional filling.
As a consequence of this, the integer effect may be qualitatively understood in terms of a non-interacting electron 
picture\cite{fqhe}. The fractional effect, on the other hand, is a much more subtle phenonemon, taking place (mostly) in 
the lowest Landau level (LLL) and possible to understand only when the interactions 
among the electrons are taken into account.
The theoretical explanation of the most prominent fractional states at $\nu = 1/m$, where $m$ is an odd integer, was given by Laughlin in 1983, when he proposed his famous many-body wave function
for the ground state\cite{laughlin83},
\be{}
\psi(z_1, z_2, ..., z_N) = \prod_{i<j} \left( z_i - z_j \right)^m e^{-\sum_i |z_i|^2/4}.
\ee
Here, $z_i \equiv (x_i + iy_i) \sqrt{\frac{eB}{\hbar c}}$ are two-dimensional complex coordinates denoting the positions of the particles in the plane. It makes intuitive sense that this wave function does a good job in minimizing the Coulomb energy of the strongly correlated electrons in the plane -- it contains the $m$th
power of a {\it Jastrow factor}, which is
a product of the 'distances' (relative coordinates) between all pairs of particles.
The Jastrow factor approaches zero when any two particles try to come close to one another and 
thus, in a sense, helps to keep the particles apart\cite{jastrowfootnote}. 
This is a useful picture to keep in mind, as this factor will show up in various contexts later in the article.

The Laughlin wave function describes a novel, inherently quantum mechanical state of matter, an {\it incompressible quantum fluid}. One of the most exotic properties of this state, and all other fractional quantum Hall states, is that it supports {\it fractional excitations}; Laughlin showed that its fundamental, charged quasiparticle excitations carry a fraction ($1/m$) of the electron charge and obey {\it fractional (anyon) statistics}\cite{leinaas77}. 
The latter means that they are neither bosons nor fermions; when two such quasiparticles are exchanged in a counterclockwise manner,
their wavefunction picks up a phase
\be{}
\psi({\bf r}_2, {\bf r}_1 ) = e^{i\pi \alpha} \psi({\bf r}_1, {\bf r}_2 )
\ee
with $\alpha = 1/m$ for the $\nu = 1/m$ state. (Bosons and fermions would correspond to $\alpha = 0$ and 1, respectively.)
Laughlin's theoretical explanation of the FQHE earned him the 1998 Nobel prize in physics, together with the experimental discoverers Tsui and St\"ormer.

As mentioned previously, a large number of FQH states have been observed since the discovery of the $\nu = 1/3$ plateau.
Most of these occur at odd denominator fractions, and many (but not all) belong to the "Jain sequences" 
$\nu = n/(2np \pm 1)$ with $n$ and $p$ integers.
There have been two main theoretical approaches to these states. In the Haldane-Halperin hierarchy 
picture\cite{haldane83,HHhierarchy}, a QH state can give rise to a sequence of 'daughter states' as successive condensates of 
quasielectrons and/or quasiholes. The basic idea
is that, once the system is in a Laughlin state and a sufficiently large number of quasielectrons or -holes have been generated 
(typically by changing the magnetic field away from its value at the center of the quantum Hall plateau), these quasiparticles
themselves may form a strongly correlated state, in much the same way as the electrons form the Laughlin state. The
result is a new incompressible QH ground state at a different filling fraction, whose quasiparticles may again condense to
form the next 'daughter' etc. 
The other approach is based on Jain's phenomenology of {\it composite fermions}\cite{jainreview}.  
The main idea of this construction is, roughly speaking, to attach an even number of vortices to each electron.
These vortices effectively cancel a part of the external magnetic field, thus mapping the electrons 
into weakly interacting {\em composite} fermions which can then be thought of as moving in a reduced magnetic field. 
This picture provides a method to construct explicit trial many-body wave functions for the ground states at 
$\nu=n/(2np\pm 1)$, as well as their quasiparticle excitations. This approach has proven highly successful, 
producing wave functions with very high overlaps with the corresponding exact states;
we will get back to the details in section \ref{sec:CF}.
Recent work\cite{CFTpapers}, based on the use of conformal field theory methods to construct hierarchical FQHE 
wave functions, illustrates that these two seemingly competing approaches are, in fact, very closely 
related\cite{read1990,blokwen}.

While almost all polarized FQH states observed to date occur at odd-denominator fractions, there is one known exception, namely, 
the gapped state at $\nu = 5/2$. It is believed to be described by the so-called {\it Pfaffian}
wave function proposed by Moore and Read\cite{MR}. The Pfaffian is the exact ground state of a three-body
repulsive interaction and describes a paired state very similar to a $p$-wave superconductor\cite{greiter92}; 
it is even more exotic than the states
discussed above, in that its quasiparticle excitations obey {\it non-Abelian} fractional statistics\cite{topQCreview}.
This generalization of 'conventional' (Abelian) anyons requires a degenerate set of $d$ states with
quasiparticles at fixed positions ${\bf r}_1$, ${\bf r}_2$, $\cdots$ ${\bf r}_n$, such that an interchange
of two quasiparticles $i$ and $j$ corresponds to a unitary operation in the subspace of these 
degenerate states,
\be{}
\psi_{\alpha} \rightarrow \rho^{(ij)}_{\alpha \beta} \psi_{\beta}.
\ee
Here, $\rho^{(ij)}$ is a $d \times d$ unitary matrix, and the set $\{ \psi_{\alpha} \}$ denotes an orthonormal basis
of the degenerate states. If the unitary matrices corresponding to different quasiparticle interchanges
do not commute, the particles are said to obey non-Abelian statistics.
The recent theoretical proposal that it might be possible to use such non-Abelian anyons in the
context of topological quantum computing\cite{topQCreview} has spurred great interest in the physics of the $\nu = 5/2$ state.
Moreover, there exist mathematical generalizations of the Pfaffian, the so-called Rezayi-Read (RR) or parafermion states\cite{RR};
although there are speculations that the recently observed QH plateau at $\nu = 12/5$\cite{xia04} might correspond to a $k=3$ parafermion
state, there is so far no unambiguous evidence of the existence of such, even more exotic, non-Abelian states in the
quantum Hall system.

Direct analogies of all the above (and many more) features of the FQHE are, in principle,
expected to occur in rapidly rotating Bose gases and have been extensively studied in
the literature in recent years. After a general introduction to rotating Bose condensates in the
next section, the
remainder of this article will be devoted to discussing these analogies
in more detail.

\section{Rotating Bose condensates}
\label{sec:RBEC}
Bose-Einstein condensation  of magnetically trapped alkali atoms was first achieved in
1995\cite{BEC1}, opening up many new directions of research on the border between
atomic and condensed matter physics. Soon after these seminal experiments, people
got interested in the rotational properties of atomic Bose condensates, and the occurrence
of quantized vortices\cite{Landau} was predicted\cite{butts}. The subsequent experimental
development has been astonishing. The first ever vortex in such an atomic cloud was 
reported by the JILA group in 1999\cite{matthews99}, and soon after by the Paris 
group\cite{madison00}. In the former, the vortex state was obtained by a direct imprinting 
of the 2$\pi$ phase shift onto the condensate, while
the latter experiment used a mechanical stirring technique, with laser beams acting basically
like a spoon in a cup of coffee\cite{stockreview}. Following these experiments, the same stirring 
technique was used to create ever larger amounts of vortices\cite{chevy00,madison1,abo1},
which could be seen to organize themselves in triangular (Abrikosov) vortex 
lattices. A well-established technique to visualize, \eg, such vortex arrays, is to perform 
absorption imaging along the rotation axis -- a picture is taken after switching off the trap and
allowing the cloud to expand for a fraction of a second. The vortices then appear as density
dips in the image, as shown in figure \ref{fig3}.
\begin{figure}
{\psfig{figure=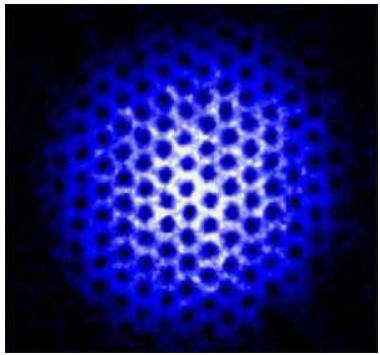, scale=0.65,angle=-0}}
\caption{Image of an Abrikosov vortex lattice in a rotating BEC.
From the JILA web page; courtesy of Eric Cornell.}
\label{fig3}
\end{figure}
There is a third method which can be
applied to further increase the angular momentum once the cloud is rotating; it is
based on "evaporative spinup", \ie evaporating atoms with angular momentum smaller than
average\cite{haljan1, engels1}. Using this technique, it has been possible to create arrays
with up to 200 vortices\cite{coddington04}, and to study detailed properties such as the vortex 
modes\cite{coddington03,schweickhard1} and vortex cores\cite{schweickhard1,coddington04}.

An interesting question to ask, then, is what will eventually happen to the vortex array if 
one keeps increasing the angular momentum of the system and thus
the density of vortices. One might expect that at some point, the vortex cores
would start to overlap, as is the case in type II superconductors about to go normal. 
However, the picture that has emerged is quite different\cite{baym1}. Since the confining
potential in typical BEC experiments is harmonic, there is a limit where the centrifugal
potential cancels the external potential and the cloud will get deconfined and fly apart
(see Sec.\ref{sec:LLL} for a more detailed discussion). 
When approaching this limit from below, the particles spread out, making
the cloud more and more pancake-shaped, and the effective interparticle interaction
becomes weaker due to the decrease in density. As we shall see in the next section, these are
preconditions for the BEC to be equivalent to a system of particles in the {\it lowest Landau
level}. The entrance into the LLL regime is signalled by a shrinking of the vortex cores
starting around the rotation frequency at which the size of a vortex core becomes comparable
to the spacing between vortices -- the ratio between the size of a vortex core and the area
occupied per vortex, saturates to a constant. This behaviour was predicted by Baym and
Pethick\cite{baym2,baym1} and confirmed in recent experiments\cite{schweickhard1}
where the transition to the LLL regime was observed to occur at around 98\% of the
deconfinement limit. Eventually, at even higher rotation, the vortex lattice is expected
to melt. This melting transition has been studied theoretically by several 
groups\cite{cooper01,sinova,baym3} and is predicted to occur around $\nu \sim 6-10$,
where the filling factor $\nu$ is the ratio of boson density to vortex density.
Beyond this point, the system enters a regime of homogeneous, strongly correlated
states of the same nature as those in the fractional quantum Hall effect.

So far, experiments have not actually reached this quantum Hall regime. Rotation
frequencies of more than 99\% of the deconfinement limit have been 
achieved\cite{schweickhard1}, which is believed to be close to the vortex 
melting transition. The main practical problem is to push the rotation further
without passing the point where the cloud flies apart. There are several very recent
proposals of ways to avoid this problem, basically by modifying the external
confinement. With these modifications, it may well be possible to reach the quantum
Hall regime with presently available experimental techniques.
We shall discuss these novel ideas in some more detail in Sec.\ref{sec:beyond}.
Meanwhile, the next three sections summarize the theoretical analogies between
rapidly rotating Bose condensates and the quantum Hall effect, simply assuming
that the system is in the lowest Landau level regime.

\section{Rotating bosons as a lowest Landau level problem}
\label{sec:LLL}
The basic insight, providing the analogy between rotating Bose condensates and the
quantum Hall system, is that the Hamiltonian of a rotating system of harmonically
confined, neutral particles is essentially equivalent to that of charged particles
in an external magnetic field. We start by making this statement more precise and
discussing under what circumstances a rotating Bose gas can be mapped to a
lowest Landau level problem. The second part of this section presents some of the
general properties of the corresponding many-body energy spectrum.
\subsection{Mapping to the LLL}
Let us consider a system of $N$ spinless bosons with mass $m$
in a harmonic trap of strength $\omega$,
rotating with angular frequency $\Omega$ 
and interacting via a short-range (delta function) potential $H_I$. In a rotating frame
the Hamiltonian can be written as
\be{rham}
H = \sum_{i=1}^N \left[\frac{\vec p_i^2}{2m} + \frac 1 2 m \omega^2 \vec r_i^2 \right]
-\Omega L_z + H_I
\ee 
where  $L_z$ denotes
the angular momentum around the rotation axis.
(For simplicity, we will set $\hbar = 1$ whenever there is no risk of confusion.)
Separating out the planar ($x,y$) part and completing the square inside the brackets, 
one can rewrite Eq.\pref{rham} as
\be{rhb}
H = \sum_{i=1}^N \left[ \frac{1}{2m}\left( \vec p_i - \vec A \right)_{\parallel}^2  
+ H_{ho}(z_i)\right] 
+(\omega -\Omega) L_z + H_I
\ee
with $\vec A = m\omega (-y,x)$, $\parallel$
denoting the planar ($x,y$) part of the Hamiltonian,
while $H_{ho}(z)$ denotes the perpendicular ($z$) part of the harmonic 
oscillator potential.
This is how the formal link
to the quantum Hall system comes about: We see that the planar part of $H$
takes the form of particles moving in an effective  "magnetic" field
$\vec B_{eff}=  \nabla \times \vec A = 2m\omega \hat z$.
The quantum mechanical one-body spectrum of this part of the Hamiltonian is given by the
so-called {\it Landau levels} (see, \eg, \cite{fqhe}), with energy
$
E_{n \parallel }= \left(n + \frac 1 2 \right) \hbar \omega_c,
$
where $n = 0, 1, 2, ...$, and $\omega_c = 2\omega$. Each Landau level is
degenerate in angular momentum, the number of states per Landau level
being equal to the number of (effective) flux quanta piercing the plane.
The single-particle wave functions in the symmetric gauge chosen here, can
be expressed as
\be{llwf}
\eta_{n,m} = N_{n,m} \, e^{-|z|^2/4} \, z^m L_n^m \left( \frac{z\bar z }{2} \right),
\ee
where $n$ is the Landau level index, $m$ denotes the angular momentum,
$N_{n,m}$ is a normalization factor, $L_n^m$ are the associated Laguerre
polynomials, and $z = \sqrt{2m\omega} (x+iy)$ is again a (dimensionless) 
complex coordinate denoting the
particle position in the plane (note the change in notation as compared to \pref{rhb}).
Now, the interaction is assumed to be weak in the sense that it does not mix different
harmonic oscillator levels. We will be interested, for a given total angular momentum, 
only in the {\em lowest lying} many-body states (the "yrast" band).
In this limit, the model may be rewritten as a lowest
Landau level (LLL) problem in the effective "magnetic" field
$B_{eff}= 2m\omega$ (and $n_z=0$ for the harmonic oscillator in the
 $z$-direction). The Hamiltonian then reduces to the form
\be{rh2}
H = (\omega - \Omega)  L + g\sum_{i<j} \delta^2({\bf r}_i - {\bf r}_j)
\ee
where $L$ denotes the total angular momentum,
$L=\sum_i l_i = L_z$.
The single particle states spanning our Hilbert space (the lowest Landau level)
are thus (omitting normalization factors)
\be{}
\eta_{0,m} =  z^m e^{-\bar z z/4}
\ee
A general bosonic many-body wave function $\psi(z_1,...z_N)$ with good angular
momentum can thus be expressed as a homogeneous, symmetric polynomial in the coordinates $\{z_i\}$, times the
exponential factor $exp(-\sum_i |z_i|^2/4)$ 
(which will be suppressed throughout most of this paper for simplicity); the degree of the
polynomial gives the total angular momentum of the state. 
\begin{figure}
{\psfig{figure=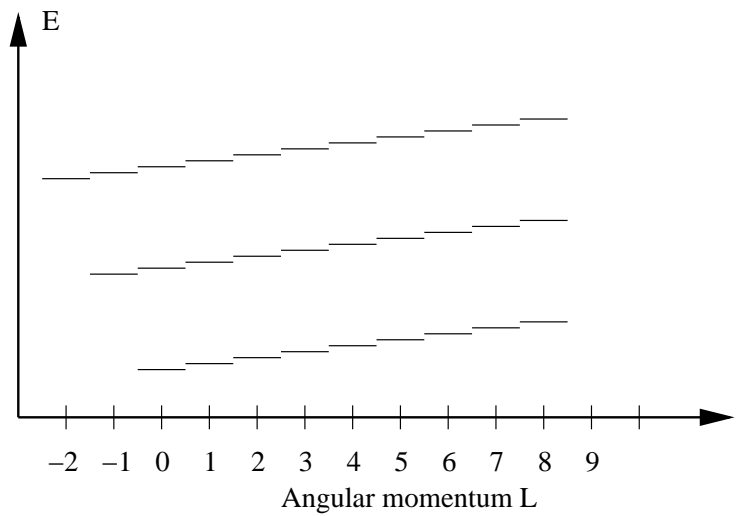, scale=0.5,angle=-0}}
\caption{Sketch of the single-particle Landau level spectrum. The slope of each level equals $\omega - \Omega$.}
\label{fig:LL}
\end{figure}

In the theoretical approach employed in the following sections, the system is simply assumed 
to be sufficiently dilute (\ie weakly interacting) to be in the lowest Landau level at all angular momenta; 
the general strategy will be to look for the lowest (interaction) energy state within this subspace 
for given $L$ and any fixed $\Omega$.
The experimental situation is somewhat different -- we saw in Sec.\ref{sec:RBEC} that
in order to become sufficiently spread out to be in this dilute regime, the cloud, in
present-day experiments, has to rotate faster than $\Omega \approx 0.98 \, \omega$.
Moreover, a more natural picture in an experimental setting 
is to think of $\Omega$ as fixed, while the system selects the ground state angular momentum 
such as to minimize $E_I + (\omega - \Omega)L$. Then, in order to avoid Landau level mixing,
the number of particles and/or $ (\omega - \Omega)$ have to be small (see fig.\ref{fig:LL}).

\subsection{Yrast spectra} \label{sec:yrast}
A convenient way of studying the many-body properties of a rotating boson system is to
display its {\it yrast spectrum}\cite{footnote2}, where the lowest many-body energy eigenvalues are
plotted as a function of total angular momentum. An example is shown in Fig. \ref{fig:yrast}, which was 
obtained by an exact diagonalization of the Hamiltonian (\ref{rham}) of the previous section for
four particles, with the lowest Landau level restriction imposed. In the absence of interactions, 
the system is highly degenerate -- the degeneracy for a given total angular momentum $L$
corresponding to the number of ways $L$ quanta of angular momentum may be distributed among
$N$ bosons. This degeneracy is lifted by the short-range repulsion, leading to the energy
band seen in the figure (the spectrum is shown for $\Omega = \omega$, \ie shows purely
the interaction energy, cfr. Eq.(\ref{rh2})). The line connecting the lowest states at different angular momenta is commonly called the yrast line.
\begin{figure}
{\psfig{figure=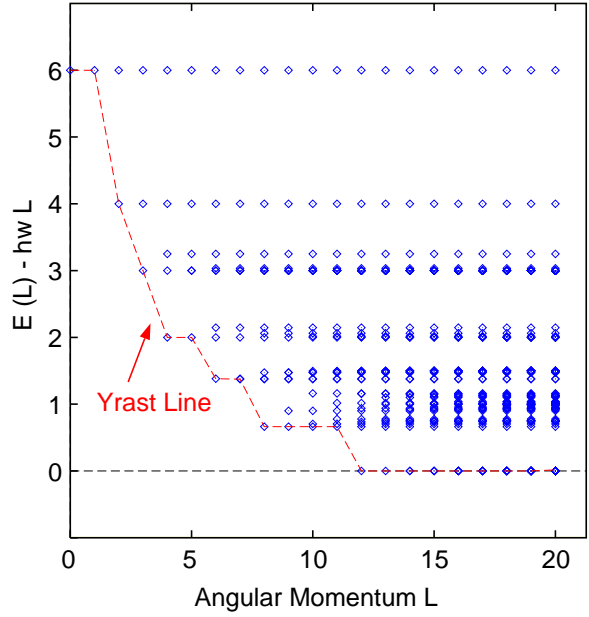, scale=0.6,angle=-0}}
\caption{Yrast spectrum, \ie many-body interaction energy as function of total angular momentum,
for four bosons in a harmonic trap with delta function repulsion. This figure was first published
in \cite{viefers1}.}
\label{fig:yrast}
\end{figure}
A number of basic properties of the system may be read out of this spectrum. First of
all, one notices that the lowest possible energy decreases with increasing angular
momentum, starting from $L=0$ where all bosons sit in the lowest angular
momentum state; this is due to the particles' ability to spread out more in the plane,
as more and more angular momentum states become available to them
(or, in other words, because the particles interact only for zero relative angular
momentum). In particular,
one notices that at $L = N(N-1)=12$ and above, the ground state has {\it zero} interaction energy. 
This is the first immediate consequence of the analogy to quantum Hall
physics: Just as in the quantum Hall effect, it is possible to construct particularly well
correlated states of the Jastrow form, \ie
\be{jb}
\psi(\{ z_i \}) = \prod_{i<j} \left( z_i - z_j \right)^{2m} \, f(z) \,
\ee
where $m$ is a positive integer, and $f(\{ z_i \})$ is some homogeneous, symmetric
polynomial in the $N$ coordinates $\{ z_i \}$. Since the repulsive interaction between the bosons 
is zero-range, any state of this form will have zero interaction energy. Since the total power
of a Jastrow factor $\prod_{i<j} \left( z_i - z_j \right)$ equals the number of pairs of particles,
$N(N-1)/2$, the Jastrow part of the wave function (\ref{jb}) contributes an angular
momentum $L_0 = mN(N-1)$. Therefore, the smallest angular momentum at which a state
of the type \pref{jb} can exist, is $N(N-1)$. At this angular momentum, the exact,
non-degenerate ground state of the system is given by the {\it Bose-Laughlin state}
\be{bl}
\psi(\{ z_i \}) = \prod_{i<j} \left( z_i - z_j \right)^{2} \, 
\ee
as was first pointed out in Ref. \cite{cooper1}. The degeneracy of the zero-energy yrast states
at $L > L_0$ corresponds to the number of ways $L - L_0$ angular momentum quanta can
be distributed among the $N$ particles; alternatively, this degeneracy can be found \cite{viefers1}
by exploiting the fact that the wave functions (\ref{jb}) describe anyons in the lowest Landau 
level\cite{hansson96} obeying Haldane's exclusion statistics\cite{haldane91} with
statistics parameter $g = 2$.

Another important feature of the yrast spectrum is that for each state, there is a set of
'daughter states' at all higher $L$, with the same (interaction) energy. These are simply
center-of-mass excitations of their 'parent' state\cite{trugman1}. At a given $L$, all
states corresponding to center-of-mass excitations of lower-$L$ eigenstates, are
orthogonal to the subspace of "new" states. According to Trugman and Kivelson\cite{trugman1},
the latter subspace consists of {\it translation invariant} (TI) polynomials, \ie
polynomials which  are invariant under a simultaneous, constant shift of all coordinates,
$z_i \rightarrow z_i + a$. A convenient basis for the TI subspace are the elementary symmetric
polynomials of degree $L$, $s_L(\{ \tilde z_i \})$ ($2 \leq L \leq N$), defined as
\be{sympol}
s_L(\tilde z_1, \dots, \tilde z_N) = {\cal S} \left\{ \tilde z_1 \tilde z_2 ... \tilde z_L \right\}.
\ee
with $\tilde z \equiv z - Z$ where $Z = \sum_i z_i / N$ is the center of mass,
and $\cal S$ denotes symmetrization of the product over all $N$ particle coordinates.
Together with $s_1(\{ z_i \}) = z_1 + z_2 + ... + z_N = NZ$, the elementary polynomials \pref{sympol} 
thus span the entire space of symmetric, homogeneous polynomials
in the lowest Landau level. We will make explicit use of the basis \pref{sympol}
when discussing low-angular momentum states in Sec.\ref{sec:lowL}.

The yrast spectra of rotating Bose condensates in the weak interaction (lowest Landau level) 
regime have been studied extensively in the literature, using a variety of methods 
including analytical studies\cite{mottelson}, mean field (Gross-Pitaevskii)
theory\cite{kavoulakis2000}, and exact numerical 
diagonalization\cite{toreblad,reimann06,manninen01}. For example, Reimann
et al\cite{reimann06} demonstrated how the presence of localized vortices in
a rotating boson cloud is revealed by periodic cusps in the yrast line of
the exact many-body spectrum.
However, the dominating line of attack in studying the analogies to quantum Hall
physics has been the use of various types of trial wave functions. 
These include bosonic versions of Laughlin- and Jain-type wave 
functions\cite{laughlin83,jainreview} and variations thereof\cite{manninen01,yannouleas},
as well as the bosonic counterparts of non-Abelian quantum Hall wave functions\cite{MR,RR}.
The results of these trial wave function studies are reviewed in more detail in the following sections.

\section{Abelian quantum Hall states}
\label{sec:aqhe}
Most known fractional quantum Hall states are Abelian. These include the Laughlin- and Jain states
and more generally all states which, in the hierarchy picture, are generated by any 
sequence of quasielectron- and quasihole condensates, as was discussed in section \ref{sec:qhe}. 
The bosonic counterparts of the Laughlin- and Jain states have been extensively studied in the literature, 
mainly numerically. Apart from exact diagonalization, a particularly widely used technique 
is the composite fermion approach; this line of work is reviewed in the first and main part of 
this section. In addition, we briefly discuss the anyonic quasiparticles of these states, the
bosonic hierarchy, and the possibility to apply the composite fermion formalism at very 
low angular momenta.
%
\subsection{The Jain sequence and composite fermions} \label{sec:CF}
%
We argued in the previous section that translation invariant (TI) states play a special
role in the rotation spectra. A particularly useful scheme of constructing TI trial wave
functions that has been widely exploited in quantum Hall physics, comes from the
phenomenology of {\em composite fermions} (CF)\cite{TIfootnote}.
Composite fermions were first introduced by Jain\cite{jainreview} and have proven
very successful
in describing FQH states, quantum dots in high magnetic fields \cite{qdcf} and, 
as we shall now discuss,
highly rotational states of Bose condensates\cite{cooper1, viefers1, chang1, regnault1}. 
In quantum Hall physics, the 
basic picture of Jain's construction is, roughly speaking, that an even number of 
vortices is bound to each electron. Each of these vortices effectively cancels one flux
quantum of the external magnetic field, and the electrons are thus
mapped into weakly interacting {\em composite}
fermions, which can be thought of as moving in a reduced magnetic field. 
Technically, "attaching a vortex"  means multiplying the wave
function by a Jastrow factor,
\be{}
\prod_{i<j} (z_i - z_j).
\ee
We already mentioned that the Jastrow factor has the effect of keeping the particles apart
-- it goes to zero if any two coordinates $z_i$ and $z_j$ approach each other.
Therefore, it takes care of much of the repulsive interaction between the
particles.
In the simplest approximation, the so-called non-interacting composite fermion 
(NICF) approach, the composite fermions are thus simply assumed to be non-interacting.
This kind of considerations led Jain to construct trial wave functions a Slater
determinant of (non-interacting) composite fermions in the reduced magnetic
field, times an even power of Jastrow factors. In the case of bosons, whose wave function
has to be symmetric rather than antisymmetric, the construction is modified by
instead binding an {\em odd} number of vortices, mapping the bosons to weakly interacting composite
fermions. In other words, bosonic trial wave functions 
with angular momentum $L$ are constructed as non-interacting fermionic
wave functions with angular momentum $L - pN(N-1)/2$, multiplied by an {\em odd}
number  $p$ of Jastrow factors, 
and projected onto the LLL,
\be{}
\psi_L = {\cal P}\, \left(  f_S(z_i, \bar z_i) \jas^p \right).
\label{cfwf}
\ee
Here, $f_S$ denotes a Slater determinant consisting of single-particle wave functions
\pref{llwf}.  The LLL
projection $\cal P$ amounts to the replacement $\bar z_i \rightarrow 2\partial/\partial z_i$ in
the polynomial part of the wave function --
the recipe is to replace all $\bar z$:s
with derivatives in the final polynomial, after multiplying out the Slater determinant
and the Jastrow factors and moving all $\bar z$:s to the left. 
It has been shown\cite{jainreview} that with this projection method, 
the single-particle wave 
functions in the CF Slater determinant may be written as
\be{}
\eta_{nl} = z^{n+l} \partial^n, ~~~~ l \geq -n
\label{etanl}
\ee
with all derivatives acting only to the right.
As this method tends
to get computationally heavy in numerical calculations with many particles and
a large number of derivatives, somewhat different methods of obtaining 
LLL wavefunctions
have been employed in most of the CF literature\cite{jainreview}. These, too,
are often referred to as projection.
Nevertheless, in this paper,
"projection" will refer to the above "brute force" procedure.
 
 Before summarizing the results obtained in the literature, 
 let us illustrate the method on two
 simple and well-known examples in the QH regime: 
 First, consider the case $L = N(N-1)$. Taking $p=1$, the
 Slater determinant $f_S$ has to contribute an angular momentum
 $N(N-1)/2$ and is
 given by putting all CFs into the lowest CF Landau level, from $l=0$ to
 $l= N-1$,
\be{}
f_S=
 \left| \begin{array}{cccc}
1 & 1 &  ... & 1 \\
z_1 & z_2 & ... & z_N \\
z_1^2 & z_2^2 & ... & z_N^2 \\
... & ... & ... & ... \\
z_1^{N-1}& z_2^{N-1} & ... & z_N^{N-1}
 \end{array} \right|
 \equiv \jas.
 \ee
 We immediately see from Eq.\pref{cfwf} that the full wave function is simply the bosonic
 Laughlin wave function \pref{bl}
 with angular momentum $L = N(N-1)$.  Next, consider the angular momentum 
 $N(N-1) - N$, corresponding to a "quasiatom" (the bosonic counterpart of a
 quasielectron) at the center. 
 While a quasihole can be seen as a local depletion of the quantum Hall liquid, a
 "quasiatom" corresponds to a local contraction, with fractional surplus charge
 (or rather particle number in our case of neutral bosons) 1/2. In the CF language, a trial wave function for such
 a quasiparticle excitation at the center, \ie with minimum angular momentum, is obtained by 
 exciting one composite fermion to the second CF Landau level, leading to the Slater
 determinant
 \be{}
f_S=
 \left| \begin{array}{cccc}
\bar z_1 & \bar z_2 & ... & \bar z_N \\
1 & 1 &  ... & 1 \\
z_1 & z_2 & ... & z_N \\
z_1^2 & z_2^2 & ... & z_N^2 \\
... & ... & ... & ... \\
z_1^{N-2}& z_2^{N-2} & ... & z_N^{N-2}
 \end{array} \right|.
 \ee
 We see that it is the excited composite fermion, with single particle wave function $\sim \bar z$
 (to be replaced by a derivative upon LLL projection) that causes the reduction of angular momentum
 as compared to the Laughlin state.
 One obtains the full trial wave function (again, apart from the exponential factor)
 \be{}
 \psi_{qp} &=& \sum_{i=1}^N (-1)^i  \partial_i \prod_{k<l ; k,l \neq i} (z_k - z_l) \prod_{m<n}^N (z_m - z_n)
 \nonumber \\
 	        &\propto& \sum_{i=1}^N \sum_{j\neq i} \frac{1}{z_i - z_j} \prod_{k\neq i} (z_i - z_k)^{-1} \, \psi_L 
 \ee
 with $\psi_L$ denoting the Laughlin state \pref{bl}.
 This wave function has very high overlap with the exact one
 ({\em \eg}, 99.7 \% for 4 bosons \cite{viefers1}). Its fermionic counterpart has been proven to 
 capture correctly both the fractional charge and the anyonic statistics of the QH
 quasielectron\cite{kjonsberg1, jeon1}, and the same can be expected to be the case
 for this bosonic version.
 
 Trial wave functions for other yrast states are constructed in similar ways. The lower
 the angular momentum, the larger the number of derivatives. Had we filled up the second
 CF Landau level, with equally many composite fermions as the first, we would have obtained a trial wave function for
 a new incompressible {\it ground state}, with filling fraction 2/3.
 In general, ground states of the {\it principal Jain sequence}  $\nu = n/(n+1)$ -- the bosonic counterpart 
 of the well-known principal Jain sequence $\nu = n/(2n+1)$ in the FQHE -- are described as $\nu^* = n$ 
 integer quantum Hall
 states of $p=1$ composite fermions, while quasihole- and "quasiatom" excitations are described by removing a CF 
from a filled CF Landau level and adding a CF to an otherwise empty CF Landau level, respectively.
Of course this construction can be generalized in the usual way by attaching a larger (odd) number
$p = 3, 5, ...$ of vortices to each boson. Note however, that for a pure delta function interaction, the 
ground states at the corresponding filling fractions $\nu = n/(np+1)$ belong to the highly degenerate subspace
of zero-energy states discussed in Sec.\ref{sec:yrast}, making this construction less relevant. Adding higher
derivatives of the delta function to the Hamiltonian (or, equivalently, using Haldane's pseudopotentials\cite{fqhe}) 
lifts this
degeneracy, and the CF construction with $p>1$ again provides good wave functions for the resulting
ground states\cite{viefers1}.

The idea of applying the CF phenomenology in the context of rotating Bose gases was first 
tested by Cooper and Wilkin\cite{cooper1} and by Viefers et al\cite{viefers1} in disk geometry\cite{fqhe}
for small numbers of particles, by comparing to exact
diagonalization results. It was shown that the approach reproduces many prominent features of the 
yrast spectrum, such as the locations of cusps in the yrast line; overlaps with the exact solutions for
a number of yrast states were computed for up to ten particles and shown to be large -- typically 99\% for five particles. 
Later, several more systematic studies were performed\cite{regnault1, chang1} in spherical geometry\cite{haldane83}.
The advantage of this theoretical approach, in which the particles move on the surface of a sphere with a radial
magnetic field produced by a magnetic monopole at the center, is that the sphere has
no boundaries. While edge effects play an important role for small systems in the plane, this geometry thus
allows for the 'simulation' of homogeneous bulk states even for the relatively modest particle numbers accessible
to numerical calculations. Instead of the complex coordinates $z_i$ discussed so far, the particle positions on
the sphere are parametrized by the polar angles ($\theta_i, \phi_i$), or more conveniently, by the spinor 
coordinates
\be{}
u_i = \cos(\theta_i/2) e^{i\phi_i/2},~~~ v_i = \sin(\theta_i/2) e^{-i\phi_i/2}.
\ee
For example, the Bose-Laughlin wave function \pref{bl} takes the form
\be{}
\psi = \prod_{i<j} \left(  u_i v_j - u_j v_i \right)^2,
\ee
and other wave functions may be translated from the plane to the sphere in a similar way.
Performing exact diagonalizations for up to 12 particles on the sphere, Regnault and Jolicoeur\cite{regnault1}
found evidence of the occurrence of incompressible (gapped) states at the principal Jain fractions
$\nu = \frac{1}{2}, \frac{2}{3}, \frac{3}{4}, \frac{4}{5}$, as well as excited states in general agreement with
the CF phenomenology.  Moreover, going away from pure delta function interaction by adding a higher-order
pseudopotential (as discussed above), they found evidence of an incompressible state at $\nu = 2/5$.
This state is not part of the principal Jain sequence; rather, it is the bosonic counterpart of the $2/7$-state
in the FQHE. Provided it is Abelian, it can be understood, in the hierarchy picture, as resulting 
from a condensate of quasiholes
in the $\nu = 1/2$ Bose-Laughlin state; in the CF picture, it belongs to the negative 
$p=3$ Jain sequence,
$\nu = n/(3n-1)$, with $n=2$. 
Although the interaction in typical experiments is dominated by $s$-wave scattering, there
exist methods to introduce and enhance an additional $d$-wave interaction\cite{marte02,thomas04,buggle04}.
This may provide a possibility, at least in principle, to observe the 2/5 state (and other states with $\nu < 1/2$).
Alternatively, one might use a system of atoms with
permanent dipolar interaction\cite{cooperPRL05} such as chromium\cite{griesmaier}.
Following up on the work of Regnault and Jolicoeur, Chang et al\cite{chang1}
performed a direct comparison between exact
diagonalization results and those predicted from the CF approach, computing energies as well as
overlaps between exact and CF wave functions for the ground states and low-lying excitations at 
$\nu = \frac 1 2, \frac 2 3$ and $\frac 3 4 $. They found that the non-interacting composite fermion approach
correctly predicts the incompressibility of the ground states at $\nu = \frac 1 2, \frac 2 3$ and $\frac 3 4 $
and produces excellent overlaps (over 97\% for up to 10 particles) for the ground state and excitations
at $\nu = \frac 1 2$ as well as the ground state at $\nu = \frac 2 3$. However, for increasing $n$, the
NICF approximation gets progressively worse, for short-range as well as Coulomb interaction, producing considerably 
poorer overlaps than in the principal Jain sequence of the electronic FQHE. In the CF language, the
interpretation is that 'residual interactions' between the composite fermions play an important role.
In particular, in the limit $n \rightarrow \infty$, \ie $\nu = 1$ (the bosonic counterpart of the metallic
$\nu = \frac 1 2$ state in the quantum Hall system), the ground state of the system can {\it not}
be described by a Fermi sea but rather appears to be a non-Abelian state, the bosonic version
of the Moore-Read Pfaffian\cite{MR}. We shall get back to this point in Sec.\ref{sec:NA}.

Additional evidence for the strongly correlated nature of the states at $\nu = 1/2$, 2/3 and 3/4
was given by Cazalilla et al\cite{cazalilla} who studied the low-energy edge excitations of
harmonically confined, rapidly rotating few-boson systems. According to Wen\cite{Wen},
the 'topological order' of a bulk quantum Hall state, implying its filling fraction, as well as 
the charge and statistics of its quasiparticle excitations, is reflected in the properties
of its edge excitations. Performing exact diagonalization studies for up to seven particles,
Cazalilla et al showed that the number of edge modes is consistent with that predicted
by Wen's theory, for  the states at $\nu = 1/2$, 2/3 and 3/4.

\subsection{Anyonic excitations}
A particularly interesting aspect of the fractional quantum Hall effect is the existence
of fractionally charged quasielectron- and quasihole excitations\cite{laughlin83}, which are 
expected to
obey fractional (anyonic) statistics\cite{leinaas77}. Obviously, the same type of
quasiparticles should occur in the bosonic quantum Hall system, with 'charge'
replaced by particle number. In the simplest case of the $\nu = 1/2$ Laughlin state,
the quasihole would be a vortex with local lack of density corresponding to
half an atom, and would obey semionic statistics, \ie quantum statistics
'halfway' between bosons and fermions. Such a quasihole, located at $z_0$,
is described by the wave function\cite{laughlin83}
\be{}
\psi_{qh}(\{ z_i \})=  \prod_i (z_i - z_0) \prod_{k<l} (z_k - z_l)^2 \, .
\ee
Paredes et al\cite{paredes} suggested that, in principle, such quasiholes
can be created by piercing the Bose-Laughlin state locally with lasers.
Adiabatically moving such a laser would then "drag" the quasihole along,
enabling controlled interchange of pairs of quasiholes and
thus a direct measurement of the anyonic phase $\pi/2$ picked up under
exchange. The latter would be particularly interesting -- despite very promising
recent experimental progress\cite{goldman} in the electronic FQH system, a 
direct and unambiguous measurement of fractional statistics is still lacking.

\subsection{Hierarchy}
In addition to the Jain states discussed above, the analogy with the FQHE in principle predicts a
large number of (Abelian) hierarchical states that do not belong to the principal Jain 
sequence\cite{haldane83,HHhierarchy,hierarchy}. An example of such a state in the 2DEG is the one recently 
observed at $\nu=4/11$\cite{pan}, whose bosonic counterpart would be $\nu = 4/7$.
Trial wave functions for these hierachy states, or at least those corresponding
to "quasiatom" (as opposed to quasihole) condensates, may be constructed using 
conformal field theory techniques\cite{CFTpapers,hierarchy,arvid}; although this 
construction is well-defined, its validity will eventually have to be determined by numerical
tests of the resulting wave functions. As discussed above, even the principal Jain states
do not describe the bosonic system as accurately as is the case in the 2DEG,
and the same may be the case for the general hierarchy construction. Moreover,
as we have seen, trial wave functions for ground states at $\nu < 1/2$ are of
interest only in systems with scattering in higher partial waves.

\subsection{Low angular momenta -- a digression} \label{sec:lowL}
Let us briefly address some interesting analytic results in a case that is very far away from 
the quantum Hall regime, namely the very lowest angular momenta
up to the single vortex, $2 \leq L \leq N$.
Within the lowest Landau level approximation, exact ground state wave functions 
for all angular momentum states in this interval were derived
 some years ago\cite{bertsch99,smith1}. They are 
given by the elementary symmetric polynomials $s_L(\tilde z_i)$ where 
$\tilde z_i = z_i - Z$ and $Z =  \sum_i z_i/N$ is the center-of-mass coordinate,
\be{eq:BP}
\psi_L^{ex} = \sum_{p_1 < p_2 < ... < p_L} (z_{p_1} - Z)(z_{p_2} - Z) \cdots (z_{p_L} - Z).
\label{TI1}
\ee 
For example, $\psi_{L=2} = {\cal S}\left[(z_{1} - Z)(z_{2} - Z)\right] $, with $\cal S$ denoting
symmetrization over all particle coordinates; for $L=N$ this expression reduces to
$\psi_{L=N} = \prod_i (z_i - Z)$. 
Since in present-day experiments, the lowest Landau
level approximation is certainly not valid at these lowest rotational states, the results 
discussed in this subsection may be somewhat academic. However, there is an interesting connection
between the exact wave functions \pref{eq:BP} and those following from a naive application
of the composite fermion construction. {\it A priori}, one would certainly expect the CF construction
to fail in this regime, at least if the usual qualitative picture of composite fermions were to be taken
literally. According to this picture, it is the flux attachment (\ie the factor $\prod (z_i - z_j)^p$) 
which makes the composite fermions weakly interacting, justifying the NICF approach.
Note, however, that since the Jastrow factor itself has an angular momentum of
$m\, N(N-1)$, the construction of CF trial wave functions at $L \sim N$ involves applying
${\cal O}(N^2)$ derivatives\cite{viefers1,korslund}. One would expect that these
derivatives acting on the Jastrow factor destroy most of the good correlations which are
at the very heart of the CF construction. It is therefore intriguing that in this regime, a naive
application of the NICF scheme produces wave functions whose overlap with the exact
ground states \pref{eq:BP} is not only large, but {\it increases} with increasing particle
number. In Ref.\cite{korslund} the single vortex state, $L=N$, was studied in detail, and
numerical calculations for up to 43 particles showed that the overlap between the CF
trial wave function and the exact analytical result \pref{eq:BP} approaches unity
for large $N$, with the difference decreasing as $\sim 1/N$ (see Fig.\ref{fig:ol}).
\begin{figure}
{\psfig{figure=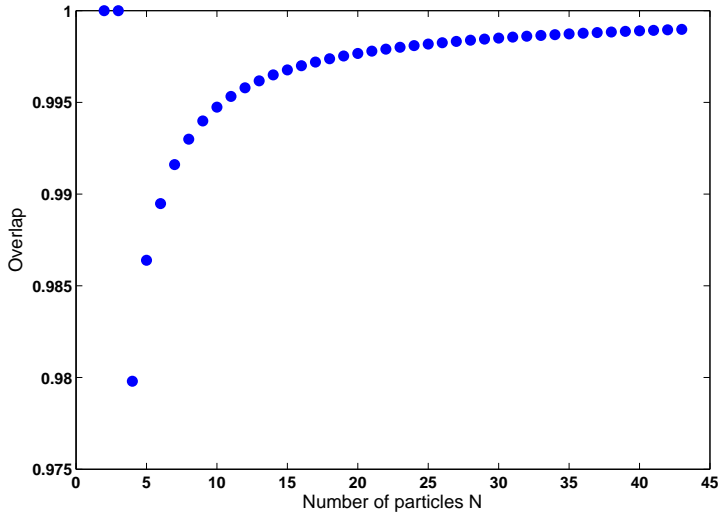, scale=0.5,angle=-0}}
\caption{Overlap between the CF trial wave function at $L=N$ and the exact one given in Eq.\pref{eq:BP},
as a function of the number of particles. Note the range of numbers on the y-axis -- the overlap equals 
99.5\% already for 10 particles.}
\label{fig:ol}
\end{figure}
Further analysis\cite{susanne} showed that this is not an artifact of the $L=N$ state.
Numerical tests up to $N=7$ for $L = N-1$ and up to
$N=8$ for $L= N-2$ show the same tendency, \ie. overlaps increasing with 
particle number. (For fixed $N$, on the other hand, overlaps tend to decrease as
$L$ decreases.)
In fact, the CF wave functions can be shown to have an analytical structure strikingly similar to
the exact ones. The simplest example is the CF state at $L=N=4$, which can be expressed as
\be{eq:LN4}
\psi_{L=N=4}^{CF} = \sum_{k=1}^4 \, \prod_{i=1}^4 \left( z_i - Z^{(k)} \right),
\ee 
with $Z^{(k)}$ denoting the 'incomplete' center-of-mass coordinate $\sum_{j \neq k} z_j/(N-1)$.
The situation becomes more complicated for higher $N$ and lower $L$, with more and more coordinates
'missing' in the center of mass (making the task of analytically proving the numerical results highly
non-trivial), but apart from this, the general structure \pref{eq:BP} is reproduced
by the CF construction.

 \section{Non-abelian quantum Hall states}\label{sec:NA}
Although the zoo of Abelian states discussed in the previous section is very rich,
it does not exhaust all possibilities of gapped quantum Hall states. The possible
existence of {\it non-Abelian} states in the quantum Hall system was already
pointed out in section \ref{sec:qhe}; in this section we shall discuss the bosonic
analogies of these states. Non-Abelian quantum Hall states have received great 
attention lately, due to the recent proposal to use their quasiparticle excitations in the
context of topological quantum computation\cite{topQCreview}. The main advantage
of this scheme is its intrinsic fault tolerance -- quantum information stored in states
with multiple non-Abelian quasiparticles is 'topologically protected', \ie immune to
local perturbations. To actually perform a quantum computation would involve
creating a state with a given number of quasiparticles at certain positions and
performing controlled braiding operations by physically dragging quasiparticles
around one another in a specified manner. One may thus speculate that, eventually,
rotating Bose condensates may provide better candidates for topological quantum
computing than the conventional QHE: In general, cold atomic systems are much easier
to control and manipulate, with a high tunability of experimental parameters, and
as already mentioned, 
there exist theoretical proposals\cite{paredes} of ways to drag quasiparticles
through the condensate. Moreover, non-Abelian states may actually be more
prominent in the bosonic case; numerical studies indicate that both the Pfaffian
and other parafermion states can be expected to occur already in the lowest Landau
level. So it is obviously of interest to study the possible occurrence of such states in the bosonic 
system. The next two subsections summarize the work that has been done in this direction. 

\subsection{The Pfaffian at $\nu = 1$}
As was pointed out in Sec.\ref{sec:CF}, $\nu = 1$ is the bosonic counterpart of the half-filled Landau
level in the electronic FQHE. In the latter, there is no quantum Hall plateau around
$\nu = 1/2$\cite{jiang89,fqhe}; rather, the system displays a compressible state, behaving like
a degenerate gas of (almost) free electrons in zero external magnetic field. This
behaviour has been explained\cite{HLR,jainreview} 
in terms of the non-interacting composite fermion model --
at $\nu = 1/2$ there are exactly two external flux quanta per electron, so after binding
two flux quanta each, the composite fermions are left in zero effective field and form
a Fermi sea. On the other hand, at $\nu = 5/2$, one does find an incompressible state
which is believed to be described by the Moore-Read wave function, the quantum Hall
analog of the paired state in a $p$-wave superconductor\cite{greiter92}; the picture is
that the presence of the two
filled lowest spin-subbands effectively modifies the interaction between the electrons
in the half-filled topmost Landau level, leading to pairing.

In the bosonic system, the situation is qualitatively different, in the sense that there 
appears to be no compressible state at $\nu = 1$. Rather, there is quite substantial
numerical evidence that the ground state corresponds to the bosonic version of the
Moore-Read state,
\be{eq:Pf}
\psi^{MR}(\{ z_i \}) &=& \prod_{i<j}(z_i - z_j)\, {\rm Pf}\left( \frac{1}{z_i - z_j} \right)
\ee
where the Pfaffian is defined as 
\be{}
 {\rm Pf}\left( \frac{1}{z_i - z_j} \right) = {\cal A} \left[ \frac{1}{(z_1 - z_2)} \,\frac{1}{(z_3 - z_4)} \cdots \frac{1}{(z_{N-1} - z_N)} \right],
\ee
with ${\cal A}$ denoting antisymmetrization over all coordinates. 
The possibility of a Pfaffian state at $\nu = 1$ was first suggested by Wilkin and Gunn\cite{wilkin2}; work by the same
group\cite{cooper01} later showed the existence of an incompressible groundstate at $\nu=1$,
and reported large overlaps (more than 96\% for six particles) between this state
and the Pfaffian trial wave function; overlap calculations by Chang et al\cite{chang1} for
up to 16 particles confirm this picture. Further numerical evidence was given by Regnault
and Jolicoeur\cite{regnault1,regnault06}, both for the ground state 
(including evidence of pairing from the form of the two-particle correlation function) 
and for quasihole excitations (correct degeneracies and high wave function overlaps).

It appears clear that the existence of the Pfaffian state at $\nu = 1$ is not an artifact of 
the short-range interaction -- on the contrary, introducing a Coulomb interaction between 
the bosons even increases the overlap\cite{chang1}. (Though the introduction of
a strong $d$-wave component in the interaction may destroy the state\cite{regnault1}).
In other words, the fact that the bosons have a stronger tendency of pairing
than their fermionic counterparts, appears to be mainly due to their quantum statistics.

\subsection{Parafermion states}
An important difference between the system at hand and the 2DEG is
that the bosonic system allows for states with $\nu > 1$ that are entirely in the
lowest Landau level, due to the absense of Pauli blocking. It is thus of interest to understand
what happens in the interval up to $\nu \approx 6$ where the system is expected to enter
the Abrikosov vortex lattice regime\cite{cooper01}. It was first suggested by Cooper
et al\cite{cooper01} that in this interval, at $\nu = k/2$; $k=3, 4, 5, ...$ one may find a 
sequence of non-Abelian incompressible states described by the parafermion
wave functions introduced by Read and Rezayi\cite{RR}.
They can be represented as\cite{cappelli}
\be{eq:RR}
\psi^{(k)}(\{ z_i \}) = {\cal S} \left[  \prod_{i<j \in A}^{N/k} (z_i - z_j)^2 \,  \prod_{k<l \in B}^{N/k} (z_k - z_l)^2 \, ...  \right],
\ee
where the system is divided into $k$ groups ($A,B,...$) each containing $N/k$ particles, 
and ${\cal S}$ denotes symmetrization over all coordinates.
The Laughlin state is recovered as the special case $k=1$, while the expression
for $k=2$ is an equivalent way of writing the Pfaffian \pref{eq:Pf}.
Generalizing the pairing in the Moore-Read state, 
these wave functions describe states with $k$-particle clustering; they are the
exact zero-energy eigenstates of a ($k+1$)-particle delta function interaction.
Performing numerical calculations on the torus, Cooper et al found large overlaps
between the Read-Rezayi states (\ref{eq:RR}) and the ground states at $\nu = k/2$
for $1 \le k \le 6$ and also recovered the correct ground state degeneracies.
Regnault and Jolicoeur\cite{regnault1,regnault06} later took these calculations
to larger systems (in spherical geometry) to see if this picture continues to hold as
one approaches the thermodynamic limit. Their results remained 
somewhat inconclusive but indicated that for a pure delta function interaction,
overlaps quickly decrease as $k$ and the number of particles are increased.
On the other hand it was shown by Rezayi et al \cite{rezayi05} that the $\nu = 3/2$
parafermion state is stabilized by introducing a moderate amount of longer-range
interaction; similar conclusions were reached for $\nu = 2$ in a very recent paper by
Cooper and Rezayi\cite{cooper07}. In principle this may be achieved in a system with
dipolar interactions or a moderate $d$-wave component. What makes these states particularly
interesting is that $k=3$ is the smallest $k$-value among the parafermion states for
which the non-Abelian statistics support universal quantum computation\cite{topQCreview}

 \section{Beyond harmonic potentials -- ways to avoid the deconfinement problem?}
 \label{sec:beyond}
 We have seen that one of the main practical obstacles to actually 
 reaching the quantum
 Hall regime experimentally, is that, for the usual harmonic confinement,
 rotation speeds exceedingly close to the deconfinement limit are required.
 The present record, with rotation at $\Omega > 0.99\, \omega$\cite{schweickhard1},
 while having reached the lowest Landau level, still lies clearly within the Abrikosov
 lattice regime. An obvious way to be able to rotate the cloud faster than $\omega$
 without it flying apart, is to modify the confining potential. This section summarizes
 a few such proposals.
 \subsection{Quartic potentials}
 Several
 theoretical studies have addressed the effect of adding a small quartic term to the 
 trap\cite{baym1}. Among the predictions for the
 vortex array regime, are the occurrence of singly quantized vortex arrays with a
 hole in the middle or, at very high rotation, a single, multiply quantized vortex
 at the center of the trap.
 In experiment, such an anharmonic trap has been created\cite{bretin04,stockreview} 
 by applying a blue detuned laser propagating along the axis of the trap. 
 This effectively amounts to adding a Gaussian potential $U(r) \sim U_0\exp(-\alpha r^2)$
 where $r$ is the (planar) distance from the axis of rotation, and the constants $U_0$ 
 and $\alpha$ are given by the parameters of the laser. For small $\alpha r^2$
 this potential is well approximated by quadratic (giving a small correction to the
 original harmonic trap) + quartic.
 
 In a very recent numerical study, Morris and Feder\cite{morris07} propose that using
 this type of quartic potential would make it possible to attain the Bose-Laughlin
 state (and other quantum Hall states) with presently accessible rotation rates.
They show that the inclusion of such a potential tends to lower the critical
rotation frequencies at which the quantum Hall states are expected to occur. 
Moreover, they predict that fine-tuning of the Gaussian parameters (depending on 
particle number and interaction strength) is necessary in order to avoid destroying the 
Laughlin state, but that the required values of these parameters are within 
experimental reach\cite{morris06}. In particular, the required experimental parameters
should become more easily accessible if the number of particles in the condensate
is reduced; this may be achieved\cite{stock05} by adding a 1D optical lattice along the axis of rotation,
splitting the condensate into an array of independent quasi-2D BECs.
 \subsection{Optical lattices}
A somewhat different, very recent theoretical proposal involving optical lattices, is
due to Bhat et al\cite{bhat07}. They suggest to include a co-rotating optical
lattice (in the tight binding regime) in addition to the harmonic potential, 
keeping the system confined  even
at critical rotation velocity $\Omega = \omega$. In addition to avoiding deconfinement of the
atom cloud, this model displays intriguing physical properties:
Mapping the system to a Bose-Hubbard model, the authors show that the rotation
introduces phase factors in the effective hopping term, 
$\hat H_{hop} \sim \hat a_i^{\dagger} \hat a_j e^{-i\phi_{ij}} \, + \, h.c.$,
where the phase depends on the rotation velocity and particle mass
 as well as the lattice spacing.
The linear response of the system to
a potential gradient (tilt of the lattice) shows quantum Hall-like features 
even for a single particle (and similarly for two particles).
The authors give the following, qualitative explanation of the analogy 
to quantum Hall physics: 
The lattice, with its tunneling
barriers, in some mean field sense mimics the repulsion experienced by a single particle from
the rest of the 2DEG, with the inaccessible regions (maxima) of the lattice
corresponding to the positions of the other electrons. Moreover, the phases
picked up by a single particle when moving around the lattice, simulate
the effect of the correlation holes (or vortices) at these 'electron positions'.
But clearly a lattice potential cannot support a liquid state,
so the exact correspondence between this system and the FQHE remains to be 
fully clarified by further studies.

Finally, it is worth mentioning that there have been other theoretical proposals 
to create a quantum Hall effect for bosonic atoms, involving {\it non-rotating} optical 
lattices \cite{palmer06},
were the magnetic field is simulated \eg by means of laser-induced hopping\cite{jaksch03} . 
A particular advantage of optical lattices
 is the extremely large degree of controllability, not only of the amount of flux per
 lattice cell but also the amount of disorder in the system.
 
 All the proposals discussed here, claim that the relevant model parameters are more or less
within reach of
 present-day experimental techniques.
 Given these latest ideas, one can certainly hope for 
 exciting experimental developments in the near future!

\section{Multicomponent Bose condensates}
\label{sec:spin}
So far we have restricted ourselves to single-component condensates of
spinless (or scalar) bosons. The quantum Hall phenomenology we have discussed is thus
analogous to the QHE of 'spinless' (fully polarized) electrons, \ie 
the case where the effective Zeeman gap is suffiently large that the spin degree of freedom
of the electrons is frozen out. Let us end with a brief discussion of the more general case where 
internal degrees of freedom, such as spin, play a role.

Polarization effects have been studied in the quantum Hall 
literature, and in particularly it has been shown that under certain conditions, the
lowest-energy charged excitation at $\nu = 1$ is a spin-textured object, a 
so-called skyrmion\cite{sondhi93}.
Moreover, there have been studies of spin polarization effects at the edge of a quantum 
Hall system, predicting the existence of spin-textured states for sufficiently smooth
confining potentials\cite{karlhede,leinaas98}. In the context of atomic Bose condensates,
there are several ways of creating systems with internal degrees of freedom, promising
an even richer phenomenology than in the quantum Hall system. 
One interesting approach to producing multicomponent Bose condensates is the simultaneous
condensation of mixtures of different atomic isotopes such as $^{85}$Rb and 
$^{87}$Rb\cite{burke98,bloch01}.
Moreover, it is possible to create {\it spinor condensates} by
trapping higher-spin atoms such as $^{87}$Rb\cite{barrett01} or 
$^{23}$Na\cite{stenger98,miesner99} in optical traps\cite{stamper98}. The advantage of 
this technique is that optical traps confine the atoms independently of their spin orientations --
as opposed to traditional magnetic traps, which typically confine only one spin 
projection, effectively giving scalar condensates\cite{barrett01,stenger98}. 

Inspired by these experimental 
advances, Reijnders et al\cite{reijnders} have performed theoretical studies of rotating spin 1
condensates in the lowest Landau level approximation, predicting a rich phase
diagram and a number of exotic states. In particular, in the quantum Hall regime,
they predict several series of novel non-Abelian states which are generalizations of the
Read-Rezayi states discussed in the previous subsection. One might expect that future
studies will continue to reveal interesting new physics in high-rotation states of
multicomponent Bose condensates.

\section{Concluding remarks}
\label{sec:concl}
A summary of the research on quantum Hall physics in rotating atomic gases is necessarily
preliminary, as the field is still highly active. At the present stage it is probably fair to
say that the theoretical side is well explored -- there are many direct analogies to the conventional
quantum Hall effect, but also physical differences, such as the expected occurrence of non-Abelian
states in the lowest Landau level. The desirable next step would be for experiments  to 'catch up'
and reach the quantum Hall regime. There is reason to be optimistic: The experimental development 
has been rapid since the first theoretical prediction of a quantum Hall effect in rotating BEC and
the first experimental creation of a quantized vortex in the late nineties. The race towards the
first experimental realization of the bosonic quantum Hall regime is going on at the time of writing,
hand in hand with new theoretical proposals how to best design such an experiment.

One can only speculate about future developments. Given the large degree of controllability 
of various parameters in cold atom experiments, one may dream of the possibility to create and
manipulate anyonic quasiparticles and directly measure their fractional statistics, Abelian or
non-Abelian. This, in turn, might be the first step towards implementing a topological quantum
computer. But this certainly cannot be expected to happen in the near future.

\vskip 4mm
\noi {\bf Acknowledgements:} 
I am very grateful to
Hans Hansson, Maria Hermanns, Jainendra Jain, Anders Karlhede, Nicolay Korslund, Thorleif Aass Kristiansen, 
Matti Manninen, and Stephanie Reimann for fruitful collaboration and discussions over the years
on various issues reviewed in this paper. 
I also wish to thank Ben Mottelson for very inspiring discussions and
for introducing us to the field of rotating Bose condensates in the first place, Murray Holland
for useful discussions on rotating optical lattices, and Jon Magne Leinaas for comments
on the manuscript.
Financial support from the Norwegian Research Council and from NordForsk is gratefully
acknowledged.

\vspace{-0.5cm}
\bibliographystyle{unsrt}

\begin{thebibliography}{10}
\bibitem{disorderfootnote} While high mobility is necessary in order for the FQHE to be observed,
the effect would actually go away if there were no impurities in the system at all. The 
occurrence of FQH plateaux requires a small but essential amount of disorder.
\bibitem{footnote1} Landau levels are the quantized energy levels of charged particles in a magnetic field.
The degeneracy $N_s$ of each Landau level equals the number of flux quanta penetrating the system, and $\nu$ is the filling fraction $N_e/N_s$, where $N_e$ is the number of electrons in the system. Some more details on Landau levels
will be discussed in section \ref{sec:LLL}.
\bibitem{jastrowfootnote} Note that a single power of the Jastrow factor would be sufficient to fulfill the
requirement that any fermionic many-body wave function has to go to zero 
as any two particles approach each other. So the Laughlin wave function goes to zero faster than what 
is simply required by the Pauli principle.
\bibitem{footnote2}"Yrast" is a term traditionally used in nuclear physics when
studying rotational spectra of nuclei. It is a Swedish word meaning "most dizzy", \ie the
highest possible angular momentum at a given energy. In the context of rotating Bose
condensates, the term 'yrast spectrum' was first introduced by Ben Mottelson\cite{mottelson}.
Similar spectra have been used extensively in the quantum Hall literature.
\end{thebibliography}

\vspace{0.5cm}
\end{document}